# Method of resolution of 3SAT in polynomial time
# By Luigi Salemi – September 2010

**Abstract**
Presentation of a Method for determining whether a problem 3Sat has solution, and if yes to find one, in time max $O(n^{15})$. Is thus proved that the problem 3Sat is fully resolved in polynomial time and therefore that it is in P, by the work of Cook and Levin, and can transform a SAT problem in a 3Sat in polynomial time (ref. Karp), it follows that P = NP. Open Source program is available at http://www.visainformatica.it/3sat

**Abstract (in Italiano)**
Presentazione di un Metodo per determinare se un problema 3Sat ha soluzione, e se si trovarne una, che richiede un tempo non superiore a $O(n^{15})$. Viene così provato che il problema 3Sat è risolto in un tempo polinomiale e quindi che lo stesso è in P, dai lavori di Cook e Levin, e dal poter trasformare un problema SAT in uno 3Sat in un tempo polinomiale (rif. Karp), ne segue che P = NP. E' disponibile, Open Source, un programma che usa il Metodo per risolvere un problema 3Sat su sito http://www.visainformatica.it/3sat

My English is bad, so this work is essential. I hope my page is enough clear, I hope someone wants to rewrite in true English.

**Introduction**
Everything comes from intuition and coincidence
Intuition: 3Sat problem is research of True Value that making TRUE all Clauses of problem. With n Variables we find $8n*(n-1)*(n-2)/6$ Clauses, this Clauses are not all in 3Sat [max 7/8 are in 3Sat]. We post in ~3Sat Clauses that not is in 3Sat. We move one Clauses from ~3Sat to 3Sat, if number of solutions not decrease we leave Clause in 3Sat else no. We end if not is possible move Clause from ~3Sat to 3Sat because we lost solutions. Now we can find one solution, n-tuple of Literal.
**This the intuition: minimize Clauses in ~3Sat.**
Coincidence: One operation move, in polynomial time, Clauses from ~3Sat to 3Sat. When it end we have in ~3Sat Clauses and in all solution triad of True Value. **This the coincidence: number of Clauses in ~3Sat is equal number of tried of True Values, also when we not have Clauses because not have tried [not have solution]**. Really funny, and I not believe in coincidence..

One example

3Sat initial [12 Clauses]: (A1 or A2 or A3) and (A1 or A2 or ~A3) and (~A1 or A2 or A3) and (~A1 or A2 or ~A3) and (~A1 or ~A2 or A3) and (~A1 or ~A2 or ~A3) and (A1 or A2 or ~A4) and (A1 or ~A2 or A4) and (~A1 or A2 or A4) and (~A1 or A2 or ~A4) and (~A1 or A3 or ~A4) and (A2 or A3 or ~A4)

~3Sat initial [20 Clauses]: (A1 or ~A2 or A3) (A1 or ~A2 or ~A3) (A1 or A2 or A4) (A1 or ~A2 or ~A4) (~A1 or ~A2 or A4) (~A1 or ~A2 or ~A4) (A1 or A3 or A4) (A1 or A3 or ~A4) (A1 or ~A3 or A4) (A1 or ~A3 or ~A4) (~A1 or A3 or A4) (~A1 or ~A3 or A4) (~A1 or ~A3 or ~A4) (A2 or A3 or A4) (A2 or ~A3 or A4) (A2 or ~A3 or ~A4) (~A2 or A3 or A4) (~A2 or A3 or ~A4) (~A2 or ~A3 or A4) (~A2 or ~A3 or ~A4)

~3Sat reduced (insert in 3Sat all Clauses possible that not lost solutions) [7 Clauses]: (~A1 or A2 or A3) (~A1 or A2 or ~A3) (~A1 or A2 or A4) (~A1 or A3 or A4) (~A1 or ~A3 or A4) (A2 or A3 or A4) (A2 or ~A3 or A4)

3Sat after move Clauses [25 Clauses]: (A1 or A2 or A3) and (A1 or A2 or ~A3) and (A1 or ~A2 or A3) and (A1 or ~A2 or ~A3) and (~A1 or ~A2 or A3) and (~A1 or ~A2 or ~A3) and (A1 or A2 or A4) and (A1 or A2 or ~A4) and (A1 or ~A2 or A4) and (A1 or ~A2 or ~A4) and (~A1 or A2 or

~A4) and (~A1 or ~A2 or A4) and (~A1 or ~A2 or ~A4) and (A1 or A3 or A4) and (A1 or A3 or ~A4) and (A1 or ~A3 or A4) and (A1 or ~A3 or ~A4) and (~A1 or A3 or ~A4) and (~A1 or ~A3 or ~A4) and (A2 or A3 or ~A4) and (A2 or ~A3 or ~A4) and (~A2 or A3 or A4) and (~A2 or A3 or ~A4) and (~A2 or ~A3 or A4) and (~A2 or ~A3 or ~A4)

Initial 3Sat with 12 Clauses and initial ~3Sat with 20 Clauses. After operation [of Method] ~3Sat have 7 Clauses and 3Sat have 25 Clauses.

Initial 3Sat with 10 Clauses and final 3Sat with 25 Clauses have 2 solutions: FTTT e FTFT, The final 3Sat is largest [with max number of Clauses] of all 3Sat with this solutions.

By FTTT we have 4 tried of True Values [index is position]: F1 T2 T3 – F1 T2 T4 – F1 T3 T4 – T2 T3 T4
By FTFT we have 4 tried of True Values: F1 T2 F3 – F1 T2 T4 – F1 F3 T4 – T2 F3 T4
Tried F1 T2 T4 is in both, then distinct tried are 7; WHAT COINCIDENCE 7 is number of Clauses in ~3Sat reduced.

Now start.

Definitions
Denote by A1, A2, .., An n Boolean Variables and by ~A1, ~A2, .., ~An their negation. Each Variable can have value "TRUE" or "FALSE", sometimes shorten the Value of Variables in "T" and "F".
We use the letters i, j, k, f, g, h, m as integer indices in the interval [1 .. n].
Denote by V1, V2, .., Vn assigning of Values to the aforementioned Variables where can be Vi = T or Vi = F.
Denote by L1, L2, .., Ln the Literal of Variables. Each Literal can be a Variable or its negation, then Li = Ai or Li = ~Ai.
Call "Pair of Literal (Li, Lj)" or simply "Pair (Li, Lj)" or even more simply if there is no ambiguity "Pair" the set of 2 literals Li, Lj with i < j .

**We are 2n\*(2n –2)/2=2n\*(n-1) possible Pairs**

Order by Pairs: (Li, Lj) < (Lk, Lh) if i < k; or i = k, j < h; or i = k, j = h, Li = Ai , Lk = ~Ai; or i = k, j = h, Li = Lk, Lj = Aj, Lh = ~Aj.

Is called "Clause" disjunction of 3 Literal [ex.: (Li or Lj or Lk)]. Clause is TRUE if is TRUE at least one Literal also is FALSE. Clause is "Described Sorted" if i < j < k.
We call "AClausola" conjunction of 3 Literal [ex.: (Li and Lj and Lk)]. AClausola is TRUE if are TRUE all Literal also is FALSE. AClausola is "Described Sorted" if i < j < k. Sometimes we write AClausola [Li Lj Lk] for save space.
Any Clause (Li or Lj or Lk) "Described Sorted" corresponds to AClausola "Described Sorted" (Li and Lj and Lk) with same Literal and vice versa.
If Ai, Aj are 2 Variables distinct [i <> j] then Ai < Aj if i < j.
We call "Tried of Variables (Ai, Aj, Ak)" o simply "Tried (Ai, Aj, Ak)" o simply "Tried" the set of 3 Variables Ai, Aj, Ak with i < j < k.
One Tried (Ai, Aj, Ak) have 8 Clause "Described Sorted" e 8 AClausola "Described Sorted" with Literal of Variables.

**We are n\*(n-1)\*(n-2)/6 Tried possible, then we are 8\*n\*(n-1)\*(n-2)/6 Clause "Described Sorted" and 8\*n\*(n-1)\*(n-2)/6 AClausole "Described Ordinate" possible**

We call "Row of Variables (Ai, Aj, Ak)" o simply "Row (Ai, Aj, Ak)" o simply "Row" the set of 0 or more AClausola "Described Sorted" all of one Tried (Ai, Aj, Aj).

**Max number of AClausola of one Row is 8**

**Max number of Row is n*(n-1)*(n-2)/6 (one for Tried)**

If Row (Ai, Aj, Ak) contains 0 AClausola is called "empty Row".

Order of Rows: Row (Ai, Aj, Ak) < Row (Af, Ag, Ah) if i < f, or i = f e j < g, or i = f, j = g e k < h.

We call "3Sat" problem to find solution at conjunction of more Clauses (ex.: (A1 or ~A2 or A3) and (A2 or A3 or ~A4) and ..) where the solution, if exists, if a set of True Value the makes TRUE any Clause, then makes TRUE formula. If set of True Value not exists then 3Sat not have solutions. We suppose that Clauses are "Described Sorted"

Now we see the Method for find solutions if exists

**Method**
We call "I3Sat" (reverse of 3Sat) set of Clauses get with substitution of any Variables with its negation (then any negation is substituted with Variable).

*Create of I3Sat have time $O(n^3)$*

Theorem 1
**3Sat have solution IFF I3Sat have solution and any solution of one is solution other with substitution T with F and F with T**
Proof
Let V1, V2, .., Vn n-tupla of True Value that resolve 3Sat and (Li or Lj or Lk) one Clause of I3Sat. Then Clause (~Li or ~Lj or ~Lk) is in 3Sat and is TRUE for Values Vi, Vj e Vk, then Values ~Vi, ~Vj and ~Vk make TRUE corresponding Clause in I3Sat. And vice versa.

Let 3Sat and Tried (Ai, Aj, Ak) we call "Complementation" to find of AClausole "Described Sorted", relatively Tried, that Clause "Described Sorted" corresponding NOT is in 3Sat.

We call "C3Sat" (Complementation 3Sat) set of n*(n-1)*(n-2)/6 order Rows, one for any Tried, where any Row have complementary AClausole [AClausola (Li and Lj and Lk) is in Row if Clause (Li or Lj or Lk) not is in 3Sat].

*Create of C3Sat have time $O(n^3)$*

n-tupla of True Values V1, V2, .., Vn solving C3Sat if make TRUE one AClausola in any Row.

**Then C3Sat with one or more empty Rows NOT have solutions**

If no n-tupla of True Values V1, V2, .., Vn solve C3Sat, then C3Sat not have solutions

Theorem 2
**3Sat have solution IFF C3Sat have solutions, from any solution of one we make solution of other simple exchange T with F and F with T .**
Proof
Let V1, V2, .., Vn one solution of 3Sat then extract Vi, Vj and Vk the Clause (Li or Lj or Lk) (Literal "L" equal negation of Variable "A" if Value "V" is TRUE or equal Variable "A" if Value "V" is FALSE) not is in 3Sat, then AClausola (Li and Lj and Lk) is in C3Sat and is TRUE by Values ~Vi, ~Vj e ~Vk. And vice versa.

**C3Sat with less n*(n-1)*(n-2)/6 AClausola not have solutions, because at least one Row is empty**

We call "CI3Sat" complementation of reverse of 3Sat

*Create CI3Sat have time $O(n^3)$, we create I3Sat and CI3Sat, but $O(n^3) + O(n^3) = O(n^3)$*

Theorem 3
**3Sat have solution IFF CI3Sat have solution and any solution of one is solution of other.**
Proof
For solution of CI3Sat we have, with inversion of True Values, solution of I3sat and, with new inversion of True Values [then we have original True Values], solution of 3Sat. And vice versa

We call "Imposition Li" the elimination, in CI3Sat, all AClausole with Literal ~Li. Then we leave, if they are, only solution where Li = TRUE.

Theorem 4
**Imposition Ai not decrease and not increase number of solution of CI3Sat where Vi = TRUE.**
Proof
To remove AClausola never increase solution.
If we have in CI3SAT solution with Vi = TRUE this make TRUE one AClausola for each Row, but in Rows with Ai and ~Ai make TRUE any AClausola with Ai [not with ~Ai]. To remove AClausole with ~Ai not decrease number of solution with Ai = TRUE.

Theorem 5
**If Imposition Ai make empty one o more Rows then we not have solution of CI3Sat with Vi = TRUE.**
Proof
We have at least one Row of CI3Sat with AClausola with only ~Ai, then none set of True Values with Ai = TRUE make TRUE one AClausola in this Row.

**Similar for Imposition ~Ai**

*Imposition Li have time $O(n^3)$*

In not empty Row of CI3Sat we have from 3 [if only one AClausola in Row] to 12 Pairs of Literal. Of the 12 possible Pairs we have some present other absent.

We call "Reduction" to find Pair of Literal (Li, Lj) absent in one Row and the remove all AClasola with this Pair in any Rows of CI3Sat.
Any Pair of Literal (Li, Lj) absent in Row [ex.: Row (Ai, Aj, Ak)] limit number of solution, that is any n-tupla of True Values that make TRUE formula (Li and Lj) not is solution of CI3Sat. In fact is missing assign of Values for Row (Ai, Aj, Ak) [in Row is missing AClasola (Li and Lj and Ak) and (Li and Lj and ~Ak)].

**Then any AClausola with Pair of Literal (Li, Lj) absent in other Row can remove from CI3Sat, this non decrease solution.**

We can execute Reduction more time [removed Pair we can make new Row without new Pair]. It end if

Case 1) at least one Row is empty.
Case 2) Reduction not remove more AClausola.

*Reduction find Pair of Literal in any Row [$O(n^3)$] and for each find AClausole with this Pair in all Rows [$O(n^3)$]. If not end we make new research many times how many are AClausole [$O(n^3)$]. Then work in time $O(n^9)$*

If CI3Sat have empty Row we call that it is empty.

Theorem 6
**Reduction not decrease and not increase number of solution of CI3Sat**
Proof
Reduction only can remove AClausola, then never increase solution
All AClausola removed have one Pair [ex.: (Li, Lj)] that is absent in other Row, than not is none n-tupla of True Values that make TRUE la formula (Li and Lj) and make TRUE one AClausola in any Row. Remove this AClausole not decrease solution.

Remove AClausola is equal to move Clause from ~3Sat to 3Sat. Reduction make this, but is sure that not lost solutions

Theorem 7
**After Reduction if Row have Literal then any Row with Variables of Literal have same Literal**
Proof
Absurd: if not then in Row (Ai, Aj, Ak) we have AClausola (Ai and Lj and Lk) and in other Row (Ai, Af, Ag) we have only AClausole (~Ai and Lf And Lg), then is missing any AClausola (Ai And Lf And Lg) for each Literal of Af and Ag. Then is missing Pair (Ai, Lf) for each Literal of Af, then is missing any AClausola type (Ai, Lf, Lj) in Row (Ai, Af, Aj), then is missing any Pair (Ai, Lj) for any Literal of Aj in any Row, also in Row (Ai, Aj, Ak). This is absurd.

**Then after Reduction if one Variable is present with both Literal, both Literal are in any Row reference this Variables**

**Trivial if Variables is absent in one Row refer Variable CI33Sat is empty.**

We call "Saturation" this operation:
- We extract all AClausole (Li and Lj and Lk) of CI3Sat
- For each AClausola we make: Imposition Ai, Imposition Aj, Imposition Ak and Reduction [for test]. If Reduction [of test] make empty CI3Sat we delete AClausola of CI3Sat and make Reduction [ultimate], else cancel Reduction [of test] and to the next AClausola.
- We Repeat 2 step previous until CI3Sat is empty [at least one Row is empty] or we cannot delete AClausole

*Saturation work in max time $O[n^{15}]$, in fact for each AClausola (max $O[n^3]$) we make 3 Imposition and 2 Reduction, then max $O[3n^3]+O[2n^9]=O[n^9]$ operation and $O[n^3]*O[n^9]=O[n^{12}]$. We repeat max $O[n^3]$ time the operation [for each AClausola] then $O[n^{15}]$.*

Theorem 8
**Saturation non decrease and not increase solution of CI3Sat**
Proof
Not increase because Saturation can only remove AClausola
Not decrease because any AClausola deleted is refer to True Values that cannot solve CI3Sat (Rif. Theorem 4), and Reduction not remove solution (Rif. Theorem 7)

From CI3Sat of n Variable we can extract CI3Sat(m) [con m < n] with AClausole with only m Variables.

Theorem 9
**Let CI3Sat Saturated of n Variable any CI3Sat(m) [con m < n] of m Variable extract from CI3Sat is Saturated**
Proof

If CI3Sat(m) not is Saturated than exist in this AClausola (Li and Lj and Lk) that Imposition Li, Imposition Lj, Imposition Lk and Reduction make empty [make empty one Row]. But then make empty same Row in CI3Sat, but CI3Sat is Saturated, then CI3Sat(m) is Saturated

Theorem 10
**If CI3Sat Saturated is empty then not have solutions**
Proof
CI3Sat have at least one Row empty, then no solutions

Theorem 11
**CI3Sat Saturated and not empty have solutions**
Proof (with building of solution)
Choice of positive Literal Ak is "Consistent Choice" if any AClausola (Ai and Aj and Ak) con i < j < k is in CI3Sat. Positive Literal not is needful, but is great simplification, However we can always get positive Literal.
We get first AClausola in CI3Sat, let (L1 and L2 and L3). If Literals are positive confirm else if someone is negative replace [in all CI3Sat] with positive [ex.: if L1 = ~A1 put A1 = ~A1 then L1 equal A1]. First choice is A1, A2, A3; choice Consistent because AClausola (A1 and A2 and A3) is in CI3Sat [with possible replace].
We Impose now A1, A2 and A3 and Reduction, we have CI3Sat_new [CI3Sat is instead initial first Imposition A1, A2, A3 and Reduction] not empty because CI3Sat è Saturated [remember: for each AClausola we can Impose of Literal of AClausola and Reduce].

Now in CI3Sat_new we are 2 type of Variables: those with one Literal e those with both Literals. If unique Literal of Variable with one Literal is negative we replace them [if Li = ~Ai put Ai = ~Ai] then unique Literal of Variables is positive. If index of Variable with one Literal is greater of index of Variable with 2 Literal exchange index [ex.: Variable Ai have only Literal Ai and Variable Ai-1 have both Literal then exchange index "i" with index "i-1" e vice versa]. Not is needful, but is great simplification

Then Variable with index <= m have only one positive Literal and Variable with index > m have both Literal. For simply we suppose that Variable with one Literal are only A1, A2, A3. Now we choice one positive Literal from any other Variable check that choice is Consistent.

We choice A4. They are (A1 and A2 and A4), (A1 and A3 and A4) and (A2 and A3 and A4) because A1, A2 and A3 are present only with positive Literal and in any Row of CI3Sat_new have any Literal of A4, Consistent choice

From Row (A1, A4, A5) we choice A5 if AClausola (A1 and A4 and A5) is present, otherwise (A1 and A4 and ~A5) is present and we choice ~A5; in last case we replace ~A5 with A5 [substitution are ALWAYS make in CI3Sat_new and CI3Sat] so we get A5. They are (A1 and A2 and A5), (A1 and A3 and A5) and (A2 and A3 and A5) because is present A5; (A1 and A4 and A5) for choice, (A2 and A4 and A5) and (A3 and A4 and A5) for presence of Pair (A4, A5) [Reduction ensure presence of one Pair of one Row in any Row of CI3Sat_new]. Consistent choice.

From Row (A1, A4, A6) we choice A6 if AClausola (A1 and A4 and A6) is present and (A1 and A4 and ~A6) is missing; we choice ~A6 if (A1 and A4 and ~A6) is present and (A1 and A4 and A6) is missing [but put ~A6 = A6 so we choice A6], if are both AClausole move to Row (A1, A5, A6) [equal reasoning] if still both AClausole move to Row (A4, A5, A6) [equal reasoning]. Finally if both AClausola choice A6.

We suppose choice from Row (A1, A4, A6) [(A1 and A4 and ~A6) is missing], check Consistence:
They are (A1 and A2 and A6), (A1 and A3 and A6) and (A2 and A3 and A6) because is present A6;

(A1 and A4 and A6) for choice. They are (A2 and A4 and A6) and (A3 and A4 and A6) because is present Pair (A4, A6)

Pair (A4, ~A6) is missing, then (A4 and A5 and ~A6) is missing, but Pair (A4, A5) is present then (A4 and A5 and A6) is present. Then they are (A1 and A5 and A6), (A2, and A5 and A6) and (A3 and A5 and A6) [because Pair (A5, A6) is present ]. Consistent choice.

We suppose choice from Row (A1, A5, A6) [(A1 and A5 and ~A6) is missing], check Consistence: They are (A1 and A2 and A6), (A1 and A3 and A6) and (A2 and A3 and A6) because A6 is present. (A1 and A4 and A6) and (A1 and A5 and A6) for choice. They are (A2 and A5 and A6) and (A3 and A5 and A6) because Pair (A5, A6) is present

Pair (A5, ~A6) is missing, then (A4 and A5 and ~A6) is missing, but Pair (A4, A5) is present then (A4 and A5 and A6) is present. Then (A2, and A5 and A6) and (A3 and A5 and A6) are present. Consistent choice.

If we choice from (A4, A5, A6) Consistence is trivial.

Now we see the general criterion for choice neatly one Literal Ak from any Variable Ak with both Literal.
  a) We see Row (A1, A4, Ak), if only one AClausola with Pair (A1, A4) get Literal Lk in this AClausola [if negative put ~Lk = Lk and get positive] if both AClausole with Pair (A1, A4) go to next step
  b) We see Row (A1, A5, Ak) and we stop if only one Literal Lk, otherwise we see Row (A1, A6, Ak) and so on to Row (A1, Ak-1, Ak), if not only one Literal go to next step
  c) We see Row (A4, A5, Ak) and we stop if only one Literal Lk, otherwise we see Row (A4, A6, Ak) and so on neatly to Row (Ak-2, Ak-1, Ak), if not only one Liter go to next step
  d) Get Ak

If Literal Lk negative put ~Lk = Lk and get positive.

For A7 is more complicated
From Row (A1, A4, A7) or (A1, A5, A7) or (A1, A6, A7) or (A4, A5, A7) or (A4, A6, A7) or (A5, A6, A7) we choice A7.

We suppose choice from Row (A1, A4, A7) [(A1 and A4 and ~A7) is missing] and check Consistence

They are (A1 and A2 and A7), (A1 and A3 and A7) and (A2 and A3 and A7) because A7 is present. (A1 and A4 and A7) for choice. They are (A2 and A4 and A7) and (A3 and A4 and A7) because Pair (A4, A7) is present

Pair (A4, ~A7) is missing , then (A4 and A5 and ~A7) and (A4 and A6 and ~A7) is missing, but Pair (A4, A5) and (A4 A6) are present [A4, A5 and A6 are Consistent] then (A4 and A5 and A7) and (A4 and A6 and A7) are present. Then Pairs (A5, A7) and (A6, A7) are present and then (A1 and A5 and A7), (A1 and A6 and A7), (A2 and A5 and A7), (A2 and A6 and A7), (A3 and A5 and A7) and (A3 and A6 and A7) are present.

To proof than (A5 and A6 and A7) is present we reason for absurd. We suppose it is missing then the CI3Sat(4) for Variables A4, A5, A6 and A7 maximum is

[A4 A5 A6] [A4 A5 ~A6] [A4 ~A5 A6] [A4 ~A5 ~A6] [~A4 A5 A6] [~A4 A5 ~A6] [~A4 ~A5 A6] [~A4 ~A5 ~A6]
[A4 A5 A7] [A4 A5 ~A7] [A4 ~A5 A7] [A4 ~A5 ~A7] [~A4 A5 A7] [~A4 A5 ~A7] [~A4 ~A5 A7] [~A4 ~A5 ~A7]
[A4 A6 A7] [A4 A6 ~A7] [A4 ~A6 A7] [A4 ~A6 ~A7] [~A4 A6 A7] [~A4 A6 ~A7] [~A4 ~A6 A7] [~A4 ~A6 ~A7]
[A5 A6 A7] [A5 A6 ~A7] [A5 ~A6 A7] [A5 ~A6 ~A7] [~A5 A6 A7] [~A5 A6 ~A7] [~A5 ~A6 A7] [~A5 ~A6 ~A7]

In Red AClausole missing: (A4 and A5 and ~A7) and (A4 and A6 and ~A7) are missing because Pair (A4, ~A7) is missing [for choice of A7] and (A5 and A6 and A7) is missing for hypothesis. **IMPORTANT (A5 and A6 and A7) not deleted by Imposition A1, A2, A3 + Reduction** because Pairs (A5, A6), (A5, A7) and (A6, A7) are present in CI3Sat_new and then not removed this AClausola with Imposition + Reduction, then this is missing in CI3Sat Saturated. But this prevent AClausola (A4 and A5 and A6) in CI3Sat because Imposition A4, A5 and A6 + Reduction put

CI3Sat(4) empty and then put empty also CI3Sat. For see we Impose AClausola [in yellow] then remain AClausole [in Green]

[A4 A5 A6] [A4 A5 ~A6] [A4 ~A5 A6] [A4 ~A5 ~A6] [~A4 A5 A6] [~A4 A5 ~A6] [~A4 ~A5 A6] [~A4 ~A5 ~A6]
[A4 A5 A7] [A4 A5 ~A7] [A4 ~A5 A7] [A4 ~A5 ~A7] [~A4 A5 A7] [~A4 A5 ~A7] [~A4 ~A5 A7] [~A4 ~A5 ~A7]
[A4 A6 A7] [A4 A6 ~A7] [A4 ~A6 A7] [A4 ~A6 ~A7] [~A4 A6 A7] [~A4 A6 ~A7] [~A4 ~A6 A7] [~A4 ~A6 ~A7]
[A5 A6 A7] [A5 A6 ~A7] [A5 ~A6 A7] [A5 ~A6 ~A7] [~A5 A6 A7] [~A5 A6 ~A7] [~A5 ~A6 A7] [~A5 ~A6 ~A7]

But this CI3Sat(4) is empty after Reduction because Variable A7 not is present with equal Literal in latest 3 Rows. (Rif. Theorem 7).
AClausola (A4 and A5 and A6) is not missing because choice A6 is Consistent, then AClausola (A5 and A6 and A7) is present and choice of A7 is Consistent.

Similarly to proof Consistent for Literal A7 choice other Row.

Focus of proof is than if we suppose absence of AClausola (A5 and A6 and A7) we have absurd [absence other AClausola than is sure present] in equal mode for any other Variable, example A8

From Row (A1, A4, A8) or (A1, A5, A8) or (A1, A6, A8) or (A1, A7, A8) or (A4, A5, A8) or (A4, A6, A8) or (A4, A7, A8) or (A5, A6, A8) or (A5, A7, A8) or (A6, A7, A8) we choice A8.
We suppose choice from Row (A1, A4, A8) [(A1 and A4 and ~A8) is missing] and check Consistence
They are (A1 and A2 and A8), (A1 and A3 and A8) and (A2 and A3 and A8) because A8 is present. (A1 and A4 and A8) for choice. They are (A2 and A4 and A8) and (A3 and A4 and A8) because Pair (A4, A8) is present
Pair (A4, ~A8) is missing, then (A4 and A5 and ~A8), (A4 and A6 and ~A8) and (A4 and A7 and ~A8) are missing, but Pairs (A4, A5), (A4, A6) and (A4, A7) are present then (A4 and A5 and A8), (A4 and A6 and A8) and (A4 and A7 and A8) are present. Then Pairs (A5, A8), (A6, A8) and (A7, A8) are present, then (A1 and A5 and A8), (A1 and A6 and A8), (A1 and A7 and A8), (A2 and A5 and A8), (A2 and A6 and A8), (A2 and A7 and A8), (A3 and A5 and A8), (A3 and A6 and A8) and (A3 and A7 and A8) are present.
To proof than (A5 and A6 and A8) is present we reason for absurd. We suppose it is missing then the CI3Sat(4) for Variables A4, A5, A6 and A8 maximum is

[A4 A5 A6] [A4 A5 ~A6] [A4 ~A5 A6] [A4 ~A5 ~A6] [~A4 A5 A6] [~A4 A5 ~A6] [~A4 ~A5 A6] [~A4 ~A5 ~A6]
[A4 A5 A8] [A4 A5 ~A8] [A4 ~A5 A8] [A4 ~A5 ~A8] [~A4 A5 A8] [~A4 A5 ~A8] [~A4 ~A5 A8] [~A4 ~A5 ~A8]
[A4 A6 A8] [A4 A6 ~A8] [A4 ~A6 A8] [A4 ~A6 ~A8] [~A4 A6 A8] [~A4 A6 ~A8] [~A4 ~A6 A8] [~A4 ~A6 ~A8]
[A5 A6 A8] [A5 A6 ~A8] [A5 ~A6 A8] [A5 ~A6 ~A8] [~A5 A6 A8] [~A5 A6 ~A8] [~A5 ~A6 A8] [~A5 ~A6 ~A8]

In Red AClausole missing: (A4 and A5 and ~A8) and (A4 and A6 and ~A8) are missing because Pair (A4, ~A8) is missing [for choice of A8] and (A5 and A6 and A8) is missing for hypothesis. **IMPORTANT (A5 and A6 and A8) not deleted by Imposition A1, A2, A3 + Reduction** because Pairs (A5, A6), (A5, A8) and (A6, A8) are present in CI3Sat_new and then not removed this AClausola with Imposition + Reduction, then this is missing in CI3Sat Saturated. But this prevent AClausola (A4 and A5 and A6) in CI3Sat because Imposition A4, A5 and A6 + Reduction put CI3Sat(4) empty and then put empty also CI3Sat. For see we Impose AClausola [in yellow] then remain AClausole [in Green]

[A4 A5 A6] [A4 A5 ~A6] [A4 ~A5 A6] [A4 ~A5 ~A6] [~A4 A5 A6] [~A4 A5 ~A6] [~A4 ~A5 A6] [~A4 ~A5 ~A6]
[A4 A5 A8] [A4 A5 ~A8] [A4 ~A5 A8] [A4 ~A5 ~A8] [~A4 A5 A8] [~A4 A5 ~A8] [~A4 ~A5 A8] [~A4 ~A5 ~A8]
[A4 A6 A8] [A4 A6 ~A8] [A4 ~A6 A8] [A4 ~A6 ~A8] [~A4 A6 A8] [~A4 A6 ~A8] [~A4 ~A6 A8] [~A4 ~A6 ~A8]
[A5 A6 A8] [A5 A6 ~A8] [A5 ~A6 A8] [A5 ~A6 ~A8] [~A5 A6 A8] [~A5 A6 ~A8] [~A5 ~A6 A8] [~A5 ~A6 ~A8]

But this CI3Sat(4) is empty after Reduction because Variable A8 not is present with equal Literal in latest 3 Rows. (Rif. Theorem 7).
AClausola (A4 and A5 and A6) is not missing because choice A6 is Consistent, then AClausola (A5 and A6 and A8) is present and choice of A8 is Consistent.

Similar to proof than (A5 and A7 and A8) is present [we suppose is missing and to proof than (A4 and A5 and A7) is absent] and (A6 and A7 and A8) [we suppose is missing and to proof than (A4 and A6 and A7) is absent]. Then choice of A8 is Consistent.

Similar for A9, A10, .., An. Then CI3Sat_new [with substituting] have solution "All TRUE" because any AClausola (Ai and Aj and Ak) with i < j < k <= n is present.

Now we remember substitution negative Literal with positive and substituting in n-tuple TRUE corresponding with FALSE. This is solution of initial CI3Sat and initial 3Sat.

In this proof number of Variable with only one Literal is insignificant [in proof we use only A1 and easy we proof Consistence for any Variable with only one Literal]. The theorem is proof in any case.

Corollary 11.1
**CI3Sat Saturated have at least one solution than make TRUE any AClausola contains**
Proof
Result of Theorem 11. We get AClausola (Li and Lj and Lk), we impose Li, Lj e Lk, we reduce and we find, for building, solution

**Group Clause of Solutions**

If 3Sat have solution (V1, V2, ..Vn) then we get n*(n-1)*(n-2)/6 Clause that are identically FALSE [ex: Vi = TRUE for any I, then any Clause (~Ai, ~Aj, ~Ak) is FALSE]. We call GCS_F(V1, V2, .., Vn) or simply GCS_F set of this Clause

Corollary 11.2
**3Sat have solution IFF exists GCS_F entirely container in ~3Sat**
Proof
If 3Sat have solution (V1, V2, .., Vn) then GCS_F(V1, V2, .., Vn) is entirely in ~3Sat. If GCS_F(V1, V2, .., Vn) is entirely in ~3Sat then (V1, V2, .., Vn) is solution of 3Sat.

CI3Sat is equivalent to ~3Sat. Then redefine GCS_F like set of AClausole identically TRUE for any solution V1, V2, .., Vn. We call GCS(V1, V2, .., Vn) or simply with GCS

Corollary 11.3
**CI3Sat ha solution IFF exist one GCS entirely container in CI3Sat**
Proof
Like previous case

Corollary 11.4
**CI3Sat Saturated container only AClausole of one GCS entirely container in CI3Sat**
Proof
For any AClausola we find one solution that put it TRUE, then exists at least one GCS container in CI3sat than container AClausola..

Corollary 11.5
**The complement of CI3Sat Saturated is largest 3Sat with all solution of initial 3Sat**
Proof
If we remove one AClausola from CI3Sat [we add one Clause to 3Sat] then at least one GCS first container now will not container, then we lost solution

Corollary 11.6

**In CI3Sat Saturated number of tried of True Values of all solution is equal number of AClausola**
Proof
For each AClausola is one tried of True Values because we can building at least one solution
For each tried of True Values in all solution is at least one solution [trivial], then al least one GCS is entirely container in CI3Sat and this container AClausola corresponding at tried
If CI3Sat is empty not have solution [zero AClausole zero tried]
WAS NOT A COINCIDENCE!

Theorem 12
**CI3Sat Saturated have solution IFF not is empty**
Proof
Theorem 10 + Theorem 11

Referenze

1) **Wikipedia**–Cook-Levin(theorem)- http://en.wikipedia.org/wiki/Cook%E2%80%93Levin_theorem
2) **Stephen Cook -**The P versus NP Problem **-** http://www.claymath.org/millennium/P_vs_NP/Official_Problem_Description.pdf
3) **Wikipedia** – P = NP Problem - http://en.wikipedia.org/wiki/P_%3D_NP_problem
4) **Wikipedia** - Karp's 21 NP-complete problems http://en.wikipedia.org/wiki/Karp's_21_NP-complete_problems